\documentstyle[12pt]{article}
\advance\voffset by -1.0cm
\advance\hoffset by -1.5cm
\input amssym.def 
\def\N{{\Bbb N}}
\def\Z{{\Bbb Z}}
\def\Q{{\Bbb Q}}
\def\R{{\Bbb R}}
\def\C{{\Bbb C}}
\def\frakg{{\frak g}}
\def\frakh{{\frak h}}
\def\Fer{\frak{Fer}}

\newcommand{\bpic}[4]{\begin{center}
                      \begin{picture}(#1,#2)(#3,#4)}
\newcommand{\epic}[1]{\end{picture}\\
                      {\small #1} \end{center}}
\newcommand{\ch}[1]{\makebox(0,0)[b]{\scriptsize$#1$}}  
\def\lch{{\large $\lhd$}}	
\def\rch{{\large $\rhd$}}

\def\SL{\mbox{SL}(2,\Z)}
\def\sign{\mbox{sign}}
\def\rank{\mbox{rank}}
\def\Res{\mbox{Res}}
\def\half{\frac{1}{2}}
\def\thalf{\frac{3}{2}}
\def\bbbone {{\mathchoice {\rm 1\mskip-4mu l} {\rm 1\mskip-4mu l}
{\rm 1\mskip-4.5mu l} {\rm 1\mskip-5mu l}}}
\def\A{{\cal A}}
\def\H{{\cal H}}

\def\Nu{{\cal N}}

\def\ket#1#2#3{\vert #1,#2,#3>}
\def\suh{\widehat{su(2)}}
\def\uh{\widehat{u(1)}}
\def\sq{\hbox{\rlap{$\sqcap$}$\sqcup$}}
\def\qed{\ifmmode\sq\else{\unskip\nobreak\hfil
\penalty50\hskip1em\null\nobreak\hfil\sq
\parfillskip=0pt\finalhyphendemerits=0\endgraf}\fi}

\newtheorem{thm}{Theorem}[section]
\newtheorem{lma}[thm]{Lemma}

\begin{document}
\thispagestyle{empty}
\begin{flushright}
DAMTP-96-06 \\
hep-th/9601163
\end{flushright}
\vspace{2.0cm}

\begin{center}

{\Large {\bf Unitarity of rational $N=2$ superconformal
\vspace*{0.5cm}

theories}} 
\vspace{2.0cm}

{\large W.\ Eholzer} and {\large M.\ R.\ Gaberdiel}\footnote{e-mail: 
W.Eholzer@damtp.cam.ac.uk and M.R.Gaberdiel@damtp.cam.ac.uk} \\ 
\vspace*{0.5cm}

{Department of Applied Mathematics and Theoretical
Physics\\
University of Cambridge, Silver Street \\
Cambridge, CB3 9EW, U.\ K.\ }
\vspace{0.5cm}

Revised Version \\
August 1996
\vspace{2.0cm}

{\bf Abstract}
\end{center}

{\leftskip=2.4truecm
\rightskip=2.4truecm

We demonstrate that all rational models of the $N=2$ super Virasoro
algebra are unitary. Our arguments are based on three different
methods: we determine Zhu's algebra $A(\H_0)$ (for which we give a
physically motivated derivation) explicitly for certain theories, we
analyse the modular properties of some of the vacuum characters, and
we use the coset realisation of the algebra in terms of $\suh$ and two
free fermions. 

Some of our arguments generalise to the Kazama-Suzuki models
indicating that all rational $N=2$ supersymmetric models might be
unitary.

}

\newpage

\section{Introduction}

Among the various conformal field theories, the supersymmetric field
theories play a special r\^ole as they are important for the
construction of realistic string theories which involve fermions.
There exist different classes of superconformal field theories which
are parametrised by $N$, the number of fermionic (Grassmann) variables
of the underlying space. For realistic string theories with $N=1$
space-time supersymmetry, the world-sheet conformal field theory is
believed to require $N=2$ supersymmetry.
\smallskip

In contrast to the $N=1$ super Virasoro algebra which is rather
similar to the non-super\-sym\-met\-ric ($N=0$) algebra, the $N=2$
algebra seems to be structurally different. For example the
Neveu-Schwarz and Ramond sector of the $N=2$ algebra are connected by
the spectral flow \cite{SS}, and the embedding structure of its Verma
modules is much more complicated \cite{Do3,Do4}. In this paper another
special feature of the $N=2$ superconformal field theory is analysed
in detail: the property that all rational theories are unitary. Here
we call a theory rational if it has only finitely many irreducible
highest weight representations, and if the highest weight space of
each of them is finite dimensional. We shall use three different
methods to analyse this problem which we briefly describe in turn.
\smallskip

It was shown by Zhu \cite{Zhu} that a theory is rational in this sense
if a certain quotient $A(\H_0)$ of the vacuum
representation $\H_0$ is finite-dimensional. This space also forms an
associative algebra, and the irreducible representations of this
algebra, the so-called Zhu algebra, are in one-to-one correspondence
with the irreducible representations of the meromorphic conformal
field theory $\H_0$. 
For the case of the $N=2$ superconformal theory, the algebra has
always the structure of a finitely generated quotient of a polynomial
algebra in two variables, and this implies that $A(\H_0)$ is 
finite dimensional for every rational theory.

In this paper we give a physically motivated
definition for $A(\H_0)$. We then show, using the embedding diagrams
of the vacuum representations of the $N=2$ algebra \cite{Do3,Do4},
that $A(\H_0)$ is infinite dimensional for a certain class of
non-unitary theories, thereby proving that these theories are not
rational. In addition, we also calculate $A(\H_0)$ explicitly for a
few special values of the central charge. We find that $A(\H_0)$ is
indeed infinite dimensional for the non-unitary cases we consider (and
finite dimensional in the unitary cases).

In order to be able to determine the dimension of $A(\H_0)$ for
arbitrary central charge one would need to know all vacuum Verma
module embedding diagrams and explicit formulae for certain singular
vectors. The embedding diagrams are known \cite{Do3,Do4}, but
sufficiently simple explicit formulae for the singular vectors do not
exist so far in general.
\smallskip

It was also shown by Zhu \cite{Zhu} that the space of torus amplitudes
(which is invariant under the modular group) is finite dimensional for
a rational superconformal field 
theory\footnote{Here we use again that for the case of the $N=2$
theory, $A(\H_0)$ is finite dimensional for every rational theory.}.
For such theories, this implies in
particular that the orbit of the vacuum character under the modular
group is a finite dimensional vector space. If this is not the case,
on the other hand, the theory cannot be rational. We determine the
vacuum characters using the embedding diagrams, and analyse the action
of the modular group on it. We then show that the relevant space is
infinite dimensional for $c\geq 3$, and for the class of non-unitary
theories with $c<3$ which was already analysed by the previous
method. As a non-trivial check, we also show that this space is finite
dimensional in the unitary minimal cases, where $c<3$. 
\smallskip

The only cases which remain can be analysed using the coset
realisation of the $N=2$ super Virasoro algebra (see {\it e.g.}\
\cite{KS})  
\begin{equation}
\label{coset}
\frac{ \suh_k \oplus (\Fer)^2}{\uh}  \,,
\nonumber
\end{equation}
which is known to preserve unitarity \cite{GKO}. Because of this
property, a non-unitary (rational) $N=2$ theory must correspond to
non-integer level for the $\suh_k$. The only remaining cases
correspond to admissible level $k\not\in\N$ for which $\suh_k$ always
has at least one (admissible) representation whose highest weight
space is infinite dimensional. Following a simple counting argument
due to Ahn {\it et al.} \cite{Ahn} we then show that this gives
rise to infinitely many inequivalent representations of the $N=2$
theory, thus proving that the theory cannot be rational. The reasoning
should be contrasted with the situation for $N=0$ and $N=1$, where the 
corresponding counting argument does not work: there the admissible
representations of $\suh_k$ give rise to the non-unitary minimal
models \cite{KacWaki,Kent}.

The last method is well amenable to generalisation. Apart from some
mathematical subtleties which we discuss, it can also be applied to
the large class of Kazama-Suzuki models, and we therefore formulate it
in this setup. As this class already provides most of the known $N=2$
models, our arguments seem to indicate that actually all rational
$N=2$ superconformal field theories might be unitary.
\smallskip

All three methods rely to varying degrees on the (conjectured)
embedding diagrams for the vacuum representations of the $N=2$ super
Virasoro algebra which we shall discuss in some detail. For example we
use the embedding diagrams to obtain a formula for the vacuum
character, whose modular properties we analyse. In the calculation of
$A(\H_0)$ we conclude from the embedding diagrams that there exist no
further relations, and finally, we check that the coset (\ref{coset})
actually realises the $N=2$ algebra by comparing the coset vacuum
character with the one obtained from the embedding diagrams.
\medskip

The paper is organised as follows. In section 2 we fix our notations
and describe the embedding diagrams for the vacuum Verma modules of
the $N=2$ super Virasoro algebra following D\"orrzapf
\cite{Do3,Do4}. In \S3 we give a physically motivated derivation for
$A(\H_0)$, and calculate it for certain cases.  In \S4, we use the
coset realisation of the algebra (\ref{coset}) to analyse the theories
which correspond to admissible level. Furthermore, we indicate how
the arguments generalise to the Kazama-Suzuki $N=2$ superconformal
theories. Finally, we remark in \S5 how the construction can be even
further generalised and give some prospective remarks.

In appendix~A we derive the vacuum character using the embedding
diagrams of \S2 as well as the coset realisation.  In appendix~B, we
analyse the modular properties of these characters, thereby showing
that certain classes cannot be rational.

\section{Preliminaries and embedding diagrams}

Let us first fix some notations and conventions. Throughout this paper
we will consider the Neveu-Schwarz sector of the $N=2$ super Virasoro 
algebra which is the infinite dimensional Lie super algebra with basis
$L_n,T_n,G^\pm_r,C$ ($n, r+\half\in\Z$) and (anti)-commutation 
relations given by 
\begin{eqnarray}
\ [L_m,  L_n]     & = & (m-n) L_{m+n}+\frac{C}{12}(m^3-m)\delta_{m+n,0}
 \nonumber \\
\ [L_m,  G^\pm_r] & = & (\frac{1}{2} m-r) G^\pm_{m+r} 
 \nonumber \\
\ [L_m,  T_n]     & = & -n T_{m+n} 
 \nonumber \\
\ [T_m,  T_n]     & = & \frac{1}{3} C m \delta_{m+n,0}  
 \nonumber \\
\ [T_m,  G^\pm_r] & = & \pm G^\pm_{m+r} 
  \nonumber\\
\ \{G^+_r,G^-_s\}  & = & 2 L_{r+s}+(r-s) T_{r+s} +
                          \frac{C}{3} (r^2-\frac{1}{4}) \delta_{r+s,0}
  \nonumber \\
\  [L_m,  C]       & = & [T_n,C] = [G^\pm_r,C] = 0
 \nonumber \\
\ \{G^+_r,G^+_s\}  & = & \{G^-_r,G^-_s\}=0 
 \nonumber
\end{eqnarray}
for all $m,n\in\Z$ and $r,s\in\Z+\frac12$.

We denote the Verma module generated from a highest weight state 
$\ket{h}{q}{c}$ with $L_0$ eigenvalue $h$, $T_0$
eigenvalue $q$ and central charge $C = c\bbbone$ by 
${\cal V}_{h,q,c}$.
An element of a highest weight representation of the 
$N=2$ super Virasoro algebra which is not proportional to the 
highest weight itself will be called a `singular vector' if it 
is annihilated by all positive modes $L_n,T_n,G^\pm_r$ 
($n,r+\half\in\N:=\{1,2\dots\}$) and is an eigenvector of $L_0$ and
$T_0$. A singular vector is called uncharged if its $T_0$ eigenvalue 
is equal to the $T_0$ eigenvalue of the highest weight state and 
charged otherwise.
The character $\chi_{\cal V}^{}$ of a highest weight representation 
$\cal V$ is defined by
$$ \chi_{\cal V}^{}(q) := q^{h-c/24} 
      \sum_{n\in \half \N -\half} \dim({\cal V}_n) q^n \, ,
$$
where ${\cal V}_n$ is the $L_0$ eigenspace with eigenvalue $h+n$,
$c$ is the central charge and $h$ the conformal dimension of
$\cal V$. The character of the Verma module ${\cal V}_{h,q,c}$ is
for example given by 
$$ \chi_{{\cal V}_{h,q,c}} = q^{h-c/24} 
       \prod_{n=1}^\infty \frac{(1+q^{n+\half})^2}{(1-q^n)^2} 
$$ 
and is called the `generic' character.

We call a meromorphic conformal field theory (MCFT) \cite{God}
rational if it possesses only finitely many irreducible MCFT
representations\footnote{ A MCFT representation is a representation
which is compatible with the vacuum representation, {\it i.e.} the
null-fields of the MCFT act trivially. We do not assume that the
graded components of a MCFT representation are finite dimensional.},
and if the highest weight space of each of them is finite
dimensional. We should stress that this definition of rationality
differs from the definition used in the mathematical literature, where
it is not assumed that the highest weight spaces are finite
dimensional, but where in addition all representations are required to
be completely reducible. It was shown by Dong {\it et al.}
\cite{DLM1} that the mathematical definition of rationality implies
the one used in this paper. On the other hand the converse is not true
as there exists a counterexample \cite{GK}.

For bosonic rational theories it has been shown by Zhu \cite{Zhu} that
the space of torus amplitudes which is invariant under the natural
action of the modular group is finite 
dimensional\footnote{Actually, Zhu showed that this is true if
$A(\H_0)$ is finite dimensional. In the case under consideration,
$A(\H_0)$ is a finitely generated quotient of the polynomial
algebra in two variables (as we shall show in section~3), 
and thus, the theory is rational if and only if $A(\H_0)$ is finite 
dimensional.}. The generalisation
to the fermionic case has been studied in \cite[Satz 1.4.6]{Hoehn}. 
If all representations are completely reducible, the space of torus
amplitudes is generated by the (finitely) many characters of the
irreducible representations. In this case, the central charge and 
the conformal dimensions of the highest weight states are all
rational numbers \cite{AM,Va}.
\smallskip

We parametrise the central charge $c$ as
$$ c(p,p') = 3(1-\frac{2p'}{p}) \, ,$$
where $p$ and $p'$ will be chosen positive for $c<3$. The well-known
series of unitary minimal models then corresponds to the central
charges being given as $c(p,1)$, where $p\ge 2$ \cite{BFK,VPYZ}.
Finally, let us denote by $\chi_{p,p'}$ the vacuum character of the
model with central charge $c(p,p')$, {\it i.e.}\ the 
character of the irreducible quotient of ${\cal V}_{0,0,c(p,p')}$.

One of the main points realised in \cite{Do2,Do3} is that there can be
up to two linearly independent uncharged singular vectors at the same
level. Indeed, this happens for example for the Verma modules related
to the unitary minimal models of the $N=2$ super Virasoro
algebra\footnote{ The embedding diagrams conjectured in
\cite{Do,Ki,Ma} for the unitary case are not correct.}. In \cite{Do3}
a complete list of all embedding diagrams of the $N=2$ super Virasoro
algebra has been conjectured.\footnote{However some of them are still
not correct \cite{Do4}.}

In contrast to the case of the Virasoro algebra it is not directly
clear how to define embedding diagrams for the Verma modules of the
$N=2$ super Virasoro algebra. This is due to the fermionic nature of
the $N=2$ algebra: suppose that there is a singular vector $\psi_{n,p}
= \theta_{n,p} \ket{h}{q}{c}$ of energy $h+n$ and charge $q+p$ in
${\cal V}_{h,q,c}$ and that ${\psi'}_{n',p'} = {\theta'}_{n',p'}
\ket{h+n}{q+p}{c}$ is  singular in ${\cal V}_{h+n,q+p,c}$. Then  
${\theta'}_{n',p'} \theta_{n,p} \ket{h}{q}{c}$ might be identically
zero in ${\cal V}_{h,q,c}$. 

Our definition of embedding diagrams of Verma modules of the $N=2$
super Virasoro algebra follows \cite{Do3} and includes only those
Verma modules which are actually embedded in the original Verma
module. To be more specific, the embedding diagram of a Verma module
of the $N=2$ super Virasoro algebra shows the highest weight vector
and all non-trivial singular vectors contained in it up to
proportionality. These vectors are connected by a line to a singular
vector if there exists an operator mapping the singular vector of
lower level (or the highest weight vector) onto the singular vector of
higher level. As in the case of the embedding diagrams of the Virasoro
algebra we shall omit lines between two vectors if these vectors are
already indirectly connected.

We also want to include in the embedding diagrams information about 
the type of the singular vectors. 
To this end we use the following notation:
the highest weight vector is denoted by a square and the singular 
vectors by circles. 
These circles are filled for singular vectors corresponding to 
Kac-determinant formula vanishings and unfilled for descendant 
singular vectors (for the explicit form of the Kac-determinant 
see \cite[eq.\ (6)]{BFK}). 
Furthermore, uncharged singular vectors which have no 
singular descendants of positive or negative charge, respectively,
are denoted by surrounding triangles pointing to the left or 
right, respectively. (These singular vectors are of type 
$\Delta(1,0)$ or $\Delta(0,1)$ in the notation of \cite{Do3}).

It has been shown in \cite{Do3} that all singular vectors
in ${\cal V}_{h,q,c}$ have charge $0$ or $\pm 1$. 
Therefore we indicate the charge of a singular vector relative to 
the highest weight vector by drawing the uncharged vectors vertically 
underneath the highest weight vector, the $-1$ charged vectors in a 
strip to the left of the highest weight vector and finally the $+1$ 
charged singular vectors in a strip to the right of the highest
weight vector. 

Let us now consider all embedding diagrams of the Verma modules 
${\cal V}_{0,0,c(p,p')}$ with $p,p'>0$.
For these values of the central charge there exist 
three types of embedding diagrams corresponding to $p=1,p'\not\in\Q$
or $p=1,p'\in\N$ or $2\le p\in\N,p'\in\N, (p,p')=1$, whose respective
embedding diagrams are shown in Fig. 2.1, Fig. 2.2 and Fig. 2.3.
\cite{Do4}\footnote{ Note that in \cite{Do3} the embedding diagram of type 
$III^0_- A^+ B^-,III^0_- A^- B^+$ (corresponding in our notation to
${\cal V}_{0,0,c(p,p')}$ with coprime $p,p'\in\N$ and $p,p'\ge2$) is
not correct. The correct embedding diagram is the same as the
embedding diagram for the case $III^0_- A^+_- B^-_+, III^0_- A^-_+
B^+_-$ (corresponding in our notation to ${\cal V}_{0,0,c(p,1)}$ with
$2\le p\in\N$)\cite{Do4}.  }.
  
\vbox{
\bpic{200}{80}{0}{0} 
\put(0,65){\ch{q:}}
\put(100,65){\ch{0}}
\put(70,65){\ch{-1}}
\put(130,65){\ch{+1}}
\put(98,50){\framebox(4,3){}}
\put(100,50){\line(-2,-1){30}}
\put(70,35){\circle*{3}}
\put(100,50){\line(2,-1){30}}
\put(130,35){\circle*{3}}
\put(100,10){\makebox(0,0)[b]{Fig. 2.1:
\small  Embedding diagram for ${\cal V}_{0,0,c(1,p')}$ with
        $p'\not\in\Q$. }}
\epic{}}

The Verma modules ${\cal V}_{0,0,c(1,p')}$ with $p'\not\in\Q$
contain only two singular vectors, namely $G^\pm_{-\half}\Omega$, where
$\Omega = \ket{0}{0}{c}$ is the vacuum vector ({\it c.f.}\ Fig. 2.1).
The Verma modules corresponding to the embedding diagram in 
Fig.\ 2.2 and Fig.\ 2.3 contain infinitely many singular vectors 
whose conformal dimensions are given by $pn(p'n \pm 1)$
($n=1,2,3,\dots$) for the uncharged singular vectors and by 
$pn(p'n+1)+p'n+\half$ and  $pn(p'n-1)-p'n-\half$ ($n=0,1,2,\dots$) 
for the charged singular vectors (for the cases corresponding to 
Fig.\ 2.2 set $p=1$). The main difference between these two diagrams
is that in Fig.\ 2.2 all singular vectors are descendants of the two
singular vectors $G^\pm_{-\half}\Omega$, which is not true in
Fig.\ 2.3.

\vbox{
\bpic{200}{200}{0}{0} 
\put(0,195){\ch{q:}}
\put(100,195){\ch{0}}
\put(70,195){\ch{-1}}
\put(130,195){\ch{+1}}
\put(98,180){\framebox(4,3){}}
\put(97,180){\line(0,-1){30}}
\put(97,150){\circle*{3}}
\put(97,150){\makebox(0,0){\lch}}
\put(103,180){\line(0,-1){30}}
\put(103,150){\circle*{3}}
\put(103,150){\makebox(0,0){\rch}}
\put(97,150){\line(0,-1){30}}
\put(97,120){\circle*{3}}
\put(97,120){\makebox(0,0){\lch}}
\put(103,150){\line(0,-1){30}}
\put(103,120){\circle*{3}}
\put(103,120){\makebox(0,0){\rch}}
\put(97,120){\line(0,-1){30}}
\put(97,90){\circle*{3}}
\put(97,90){\makebox(0,0){\lch}}
\put(103,120){\line(0,-1){30}}
\put(103,90){\circle*{3}}
\put(103,90){\makebox(0,0){\rch}}
\put(97,90){\line(0,-1){30}}
\put(97,60){\circle*{3}}
\put(97,60){\makebox(0,0){\lch}}
\put(103,90){\line(0,-1){30}}
\put(103,60){\circle*{3}}
\put(103,60){\makebox(0,0){\rch}}
\put(97,50){\makebox(0,0){$\vdots$}}
\put(103,50){\makebox(0,0){$\vdots$}}
\put(100,180){\line(-2,-1){30}}
\put(70,165){\circle*{3}}
\put(70,164){\line(0,-1){28}}
\put(70,135){\circle{3}}
\put(70,134){\line(0,-1){28}}
\put(70,105){\circle{3}}
\put(70,104){\line(0,-1){28}}
\put(70,75){\circle{3}}
\put(70,74){\line(0,-1){28}}
\put(70,45){\circle*{3}}
\put(97,151){\line(-2,1){27}}
\put(97,149){\line(-2,-1){27}}
\put(97,121){\line(-2,1){27}}
\put(97,119){\line(-2,-1){27}}
\put(97,91){\line(-2,1){27}}
\put(97,89){\line(-2,-1){27}}
\put(97,61){\line(-2,1){27}}
\put(97,59){\line(-2,-1){27}}
\put(70,35){\makebox(0,0){$\vdots$}}
\put(100,180){\line(2,-1){30}}
\put(130,165){\circle*{3}}
\put(130,164){\line(0,-1){28}}
\put(130,135){\circle{3}}
\put(130,134){\line(0,-1){28}}
\put(130,105){\circle{3}}
\put(130,104){\line(0,-1){28}}
\put(130,75){\circle{3}}
\put(130,74){\line(0,-1){28}}
\put(130,45){\circle*{3}}
\put(103,151){\line(2,1){26}}
\put(103,149){\line(2,-1){26}}
\put(103,121){\line(2,1){26}}
\put(103,119){\line(2,-1){26}}
\put(103,91){\line(2,1){26}}
\put(103,89){\line(2,-1){26}}
\put(103,61){\line(2,1){26}}
\put(103,59){\line(2,-1){26}}
\put(130,35){\makebox(0,0){$\vdots$}}
\put(100,10){\makebox(0,0)[b]{Fig. 2.2:
\small  Embedding diagram for ${\cal V}_{0,0,c(1,p')}$ with
        $p'\in\N$. }}
\epic{}}

\vbox{
\bpic{200}{180}{0}{0} 
\put(0,195){\ch{q:}}
\put(100,195){\ch{0}}
\put(50,195){\ch{-1}}
\put(150,195){\ch{+1}}
\put(98,180){\framebox(4,3){}}
\put(100,180){\line(0,-1){30}}
\put(100,150){\circle*{3}}
\put(100,150){\line(0,-1){30}}
\put(97,120){\circle*{3}}
\put(97,120){\makebox(0,0){\lch}}
\put(103,120){\circle*{3}}
\put(103,120){\makebox(0,0){\rch}}
\put(97,120){\line(0,-1){30}}
\put(103,120){\line(0,-1){30}}
\put(97,90){\circle*{3}}
\put(103,90){\circle*{3}}
\put(97,90){\makebox(0,0){\lch}}
\put(103,90){\makebox(0,0){\rch}}
\put(97,90){\line(0,-1){30}}
\put(103,90){\line(0,-1){30}}
\put(97,60){\circle*{3}}
\put(103,60){\circle*{3}}
\put(97,60){\makebox(0,0){\lch}}
\put(103,60){\makebox(0,0){\rch}}
\put(97,60){\line(0,-1){30}}
\put(103,60){\line(0,-1){30}}
\put(97,30){\circle*{3}}
\put(103,30){\circle*{3}}
\put(97,30){\makebox(0,0){\lch}}
\put(103,30){\makebox(0,0){\rch}}
\put(50,160){\circle*{3}}
\put(150,160){\circle*{3}}
\put(50,159){\line(0,-1){28}}
\put(150,159){\line(0,-1){28}}
\put(50,130){\circle{3}}
\put(150,130){\circle{3}}
\put(50,129){\line(0,-1){28}}
\put(150,129){\line(0,-1){28}}
\put(50,100){\circle{3}}
\put(150,100){\circle{3}}
\put(50,99){\line(0,-1){28}}
\put(150,99){\line(0,-1){28}}
\put(50,70){\circle{3}}
\put(150,70){\circle{3}}
\put(50,69){\line(0,-1){28}}
\put(150,69){\line(0,-1){28}}
\put(50,40){\circle{3}}
\put(150,40){\circle{3}}
\put(100,180){\line(5,-2){49}}
\put(100,180){\line(-5,-2){49}}
\put(100,150){\line(5,-2){49}}
\put(100,150){\line(-5,-2){49}}
\put(103,118){\line(5,-2){46}}
\put(97,118){\line(-5,-2){46}}
\put(103,88){\line(5,-2){46}}
\put(97,88){\line(-5,-2){46}}
\put(103,58){\line(5,-2){46}}
\put(97,58){\line(-5,-2){46}}
\put(103,122){\line(5,4){46}}
\put(97,122){\line(-5,4){46}}
\put(103,92){\line(5,4){46}}
\put(97,92){\line(-5,4){46}}
\put(103,62){\line(5,4){46}}
\put(97,62){\line(-5,4){46}}
\put(103,32){\line(5,4){46}}
\put(97,32){\line(-5,4){46}}
\put(97,20){\makebox(0,0){$\vdots$}}
\put(103,20){\makebox(0,0){$\vdots$}}
\put(50,30){\makebox(0,0){$\vdots$}}
\put(150,30){\makebox(0,0){$\vdots$}}
\put(100,0){ \makebox(0,0)[b]{Fig. 2.3: 
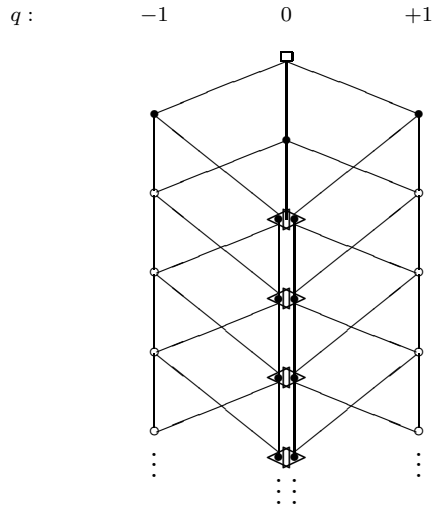
\small  Embedding diagram for ${\cal V}_{0,0,c(p,p')}$ with
        coprime $p,p'$ and  $p\ge2$.}}
\epic{}}

The line from the first uncharged singular vector to the two
dimensional space of singular vectors at level $p'(p+1)$ in Fig.\ 2.3 
deserves comment: it means that the first uncharged singular vector
at level $p'(p-1)$ has exactly one descendent uncharged singular 
vector at level $p'(p+1)$ which is contained in the two-dimensional
space of singular vectors at this level. The module generated from
this descendent singular vector contains singular vectors both of
charge $0$  and  $\pm 1$.  

Let us end with two comments on the embedding diagrams of the vacuum
Verma modules for $c\ge3$ which we will need in the 
appendix (for details see \cite{Do3}). 
For $c=3$ there are infinitely many singular vectors which are all 
embedded in the two generic singular vectors $G^{\pm}_{-\half}\Omega$.
For $c>3$ all embedding diagrams terminate, {\it i.e.}\ there are only
finitely many singular vectors contained in the vacuum Verma modules.

\section{Zhu's algebra}

In this section we shall first give a physically motivated derivation
of Zhu's algebra; we shall then use this formulation to determine
$A(\H_0)$ for a certain class of theories of the $N=2$ super Virasoro
algebra, and  for some special cases.
\smallskip

A conceptually interesting way to determine all irreducible
representations of a (bosonic) meromorphic conformal field theory is
the method introduced by Zhu \cite{Zhu}, whereby one associates an
associative algebra, usually denoted by $A(\H_0)$, to the vacuum
representation $\H_0$ of a conformal field theory. It was shown by Zhu
that the irreducible representations of this associative algebra are
in one-to-one correspondence with the irreducible representations of
the meromorphic field theory $\H_0$. To define this algebra, a certain
product structure was introduced by means of some rather complicated
formulae, and it was not clear, how this construction could be
understood from the more traditional point of view of conformal field
theory. Here we shall give a different derivation for this 
algebra, from which it will be immediate that all representations of
$\H_0$ have to be representations of $A(\H_0)$ (this derivation
follows in spirit \cite{FeiFu} and \cite{Watts}); to show the converse
direction, a similar argument as the one given in \cite{Zhu} would be 
sufficient. Another virtue of our derivation is that the Neveu-Schwarz
fermionic case (which has by now been independently worked out by Kac
and Wang in \cite{KacWang}) can essentially be treated on the same
footing.

To fix notation, let us denote the modes of a holomorphic field $S(z)$
of conformal weight $h$ by
\begin{equation}
S(z) = \sum_{l\in\Z} S_{-l} \; z^{l-h} \,. \nonumber
\end{equation}

Given two representations of the chiral symmetry algebra ${\cal A}$, 
${\cal H}_1$ and ${\cal H}_2$,
and two points $z_1, z_2 \in \C$ in the complex plane, 
the fusion tensor product can be defined by
the following construction \cite{Gab93}. First we consider the product
space $\left( {\cal H}_1 \otimes {\cal H}_2 \right)$ on which two
different actions of the chiral algebra are given by the two
comultiplication formulae \cite{Gab94}
\begin{eqnarray}
{\displaystyle \Delta_{z_1,z_2}(S_{n}) = \widetilde{\Delta}_{z_1,z_2}(S_{n})}
= &&
{\displaystyle \sum_{m=1-h}^{n} \left( \begin{array}{c} n+h-1 \\ m+h-1
\end{array} \right)
z_1^{n-m} \left(S_{m} \otimes \bbbone\right)  }
\hspace*{3.3cm} \nonumber \\
\label{chir1}
&&  \hspace*{1.0cm} {\displaystyle +\, \varepsilon_{1}
\sum_{l=1-h}^{n} \left( \begin{array}{c} n+h-1 \\ l+h-1
\end{array} \right)
z_2^{n-l} \left(\bbbone \otimes S_{l} \right)\,,}  
\end{eqnarray}
\begin{eqnarray}
{\displaystyle \Delta_{z_1,z_2}(S_{-n})} = &&
{\displaystyle \sum_{m=1-h}^{\infty} \left( \begin{array}{c} n+m-1 \\ n-h
\end{array} \right) (-1)^{m+h-1}
z_1^{-(n+m)} \left(S_{m} \otimes \bbbone \right) }
\hspace*{3.0cm} \nonumber \\
\label{chir2}
&&  \hspace*{3.0cm} {\displaystyle +\, \varepsilon_{1}
\sum_{l=n}^{\infty} \left( \begin{array}{c} l-h \\ n-h
\end{array} \right)
(-z_2)^{l-n} \left(\bbbone \otimes S_{-l}  \right)\,,} \\
\label{chir2'}
{\displaystyle \widetilde{\Delta}_{z_1,z_2}(S_{-n})} = &&
{\displaystyle 
\sum_{m=n}^{\infty} \left( \begin{array}{c} m-h \\ n-h
\end{array} \right) 
(-z_1)^{m-n} \left(S_{-m} \otimes \bbbone \right) }
\hspace*{4.5cm} \nonumber \\
&& \hspace*{1.0cm} {\displaystyle + \varepsilon_{1}
\sum_{l=1-h}^{\infty} \left( \begin{array}{c} n+l-1 \\ n-h
\end{array} \right) (-1)^{l+h-1}
z_2^{-(n+l)} \left(\bbbone \otimes S_{l}  \right)\,,}
\end{eqnarray}
where in (\ref{chir1}) we have $n\geq 1-h$, in
(\ref{chir2},\,\ref{chir2'}) $n\geq h$, and $\varepsilon_{1}$ is $\mp 1$
according to whether the left-hand vector in the tensor product
and the field $S$ are both fermionic or not.\footnote{The second
formula differs from the one given in \cite{Gab94} by a different
$\varepsilon$ factor. There the two comultiplication formulae were
evaluated on different branches; this is corrected here.} 

The fusion tensor product is then defined as the quotient of the
product space by all relations which come from the equality of
$\Delta_{z_1,z_2}$ and $\widetilde{\Delta}_{z_1,z_2}$
\begin{equation}
\left( {\cal H}_1 \otimes {\cal H}_2 \right)_f :=
\left( {\cal H}_1 \otimes {\cal H}_2 \right) / 
(\Delta_{z_1,z_2} - \widetilde{\Delta}_{z_1,z_2}) \,. \nonumber
\end{equation}
It has been shown for a number of examples that this definition
reproduces the known restrictions for the fusion rules
\cite{Gab93,Gab94}.

To analyse the possible representations of the meromorphic field
theory $\H_0$, let us consider the fusion product of a
given representation $\H$ at $z_2=0$ with the vacuum
representation $\H_0$ at $z_1=z$. We shall be interested in the
quotient of the fusion product by all states of the form
\begin{equation}
\Delta_{z,0} ( \A_{-} ) ( \H_0 \otimes \H )_f \,, \nonumber
\end{equation}
where $\A_{-}$ is the algebra generated by all negative
modes. (In the conventional approach to fusion in terms of $3$-point
functions, all such states vanish if there is a highest weight vector
at infinity.) Using the comultiplication $\Delta_{z,0}$, it is clear
that we can identify this quotient space with a certain subspace of
\begin{equation}
\left(\H_0 \otimes \H \right)_f / \Delta_{z,0} (\A_{-}) \left( \H_0 \otimes
\H \right)_{f} \subset \left( \H_0 \otimes \H^{(0)} \right)\,, \nonumber
\end{equation}
where $\H^{(0)}$ is the highest weight space of the representation
$\H$. The idea is now to analyse this quotient space for the universal
highest weight representation $\H=\H_{univ}$, {\it i.e.} to use no
property of $\psi\in\H^{(0)}_{univ}$, other than that it is a highest
weight state. We can then identify this quotient space with a certain
quotient of the vacuum representation $\H_0$, thus defining $A(\H_0)$,
\begin{equation}
\left( A(\H_0) \otimes \H^{(0)}_{univ} \right) = 
\left(\H_0 \otimes \H_{univ} \right)_f / 
\Delta_{z,0} (\A_{-}) \left( \H_0 \otimes \H_{univ} \right)_f  \,.
\end{equation}

In order to do this analysis without using any information about
$\psi$, we have to find a formula for
\begin{equation}
(\bbbone\otimes S_0) (\H_0 \otimes \psi) \hspace*{1cm} 
\mbox{mod} \;\;\;\; \Delta_{z,0} (\A_{-}) (\H_0 \otimes \H_{univ}) \,,
\nonumber 
\end{equation}
in terms of modes acting on the left-hand factor in the tensor
product, where $S$ is any bosonic field, {\it i.e.}\ $S$ has integral
conformal dimension $h$. 

The crucial ingredient we shall be using is the observation
\begin{equation}
\widetilde{\Delta}_{0,-z} (S_{-h}) = \Delta_{z,0} ( e^{z L_{-1}}
S_{-h} e^{-z L_{-1}} ) \in \Delta_{z,0} (\A_{-}) \,. \nonumber
\end{equation}
Hence we have (for $h\geq 2$)
\begin{eqnarray}
0 & \cong & \widetilde{\Delta}_{0,-z}(S_{-h}) + 
\sum_{l=1-h}^{-1} z^{-(h+l)} \Delta_{z,0} (S_l) \nonumber \\
& = &
\left( S_{-h} \otimes \bbbone \right) + 
\sum_{l=1-h}^{\infty} 
\left( \begin{array}{c} h+l-1 \\ h-h \end{array} \right) (-1)^{l+h-1}
(-z)^{-(h+l)} \left(\bbbone\otimes S_l \right) \nonumber \\
& & 
 + \sum_{l=1-h}^{-1} \left\{ z^{-(h+l)} \left( \bbbone\otimes S_l \right)
+  z^{-(h+l)} \sum_{m=1-h}^{l} 
\left( \begin{array}{c} l+h-1 \\ m+h-1 \end{array} \right)
z^{l-m} \left( S_m \otimes \bbbone \right)\right\} \nonumber \\
& = &
\left( S_{-h} \otimes \bbbone \right) 
- z^{-h} \left(\bbbone \otimes S_0 \right) 
+ \left(\bbbone \otimes {\cal A}_{+} \right)  \nonumber \\
& &  
+ \sum_{l=1-h}^{-1} \; \sum_{m=1-h}^{l} z^{-(h+m)}
\left( \begin{array}{c} l+h-1 \\ m+h-1 \end{array} \right)
\left(S_m \otimes \bbbone \right) \,, \nonumber
\end{eqnarray}
where ${\cal A}_{+}$ is the algebra generated by the positive
modes, and $\cong$ denotes equality up to terms in the
quotient. Evaluated on $(\H_0\otimes \psi)$, where $\psi$ is a
highest weight, we then have 
\begin{equation}
\label{zero}
\left(\bbbone \otimes S_0 \right) \cong
z^{h} \left(S_{-h}\otimes\bbbone\right) 
+ \sum_{l=1-h}^{-1} \; \sum_{m=1-h}^{l} z^{-m}
\left( \begin{array}{c} l+h-1 \\ m+h-1 \end{array} \right)
\left(S_m \otimes \bbbone \right) \,.
\end{equation}

In particular, we can use this result to obtain a formula for the
action of $\Delta_{z,0}(S_0)$ on $(\psi^0 \otimes \psi)$ modulo
vectors in the quotient, where $\psi^0\in A(\H_0)$. (It is clear that
$\Delta_{z,0}(S_0)$ is well-defined on the quotient.) We calculate
\begin{eqnarray}
\Delta_{z,0}(S_0) & \cong &
z^{h} (S_{-h} \otimes \bbbone) + (S_0 \otimes \bbbone)
\nonumber \\
& & 
+ \sum_{m=1-h}^{-1} z^{-m} (S_m \otimes \bbbone) 
\left\{
\left( \begin{array}{c} h-1 \\ m+h-1 \end{array} \right)
+ \sum_{l=m}^{-1} 
\left( \begin{array}{c} l+h-1 \\ m+h-1 \end{array} \right) \right\}
\nonumber \\
& = &
\sum_{m=0}^{h} z^m 
\left( \begin{array}{c} h \\ m \end{array} \right) 
(S_{-m} \otimes \bbbone)\,, \nonumber
\end{eqnarray}
where we have used the identity 
\begin{equation}
 \sum_{k=0}^{l} \left( \begin{array}{c} a+k \\ k \end{array} \right)
= \left( \begin{array}{c} a+l+1 \\ l \end{array} \right)
 \label{binid}
\end{equation}
(see {\it e.g.}\ \cite[p.\ 174]{Formula}) to rewrite the sum in
curly brackets. This reproduces precisely the product formula of Zhu 
for $z=1$ \cite{Zhu}  
\begin{equation}
S \star \psi^0 = \sum_{m=0}^{h} 
\left( \begin{array}{c} h \\ m \end{array} \right)
S_{m-h} \psi^0 \,. \nonumber
\end{equation}
If $S$ is the field corresponding to a state in the subspace
of the vacuum representation $\H_0$ by which we quotient to obtain
$A(\H_0)$, then its zero mode vanishes by definition on all highest
weight states. This implies that the product structure defined by
the action of $\Delta_{z,0}(S_0)$ gives rise to a well-defined product 
on $A(\H_0)$. 

We have now achieved our first goal, namely to express the zero modes
of the holomorphic fields on a highest weight state in terms of modes
acting in the vacuum representation, modulo terms which vanish if
there is a highest weight vector at infinity. In the next step we want
to derive the relations by which the vacuum representation has to be
divided in order to give $A(\H_0)$. In particular, we shall see that we
can express all states of the form $(S_{-n}\otimes \bbbone) ({\cal
H}_0 \otimes \psi)$ with $n>h$ by corresponding states with $n\leq
h$. (Again this can be done without using any property of $\psi$
other than that it is a highest weight vector.)

We shall do the calculation for the bosonic case first, 
{\it i.e.}\ for $h\in\N$; we shall explain later, what modifications
arise in the fermionic case. As before we have
\begin{eqnarray}
0 & \cong & 
\widetilde{\Delta}_{0,-z} (S_{-n}) 
+ \sum_{l=1-h}^{-1}
\left( \begin{array}{c} n+l-1 \\ n-h \end{array} \right)
(-1)^{h-n} z^{-(n+l)} \Delta_{z,0}(S_l) \nonumber \\
& = & (S_{-n}\otimes \bbbone) -
\left( \begin{array}{c} n-1 \\ n-h \end{array} \right) (-1)^{h-n}
z^{-n} (\bbbone\otimes S_0) + (\bbbone\otimes \A_{+}) \nonumber \\
& & 
+ \sum_{l=1-h}^{-1}
\left( \begin{array}{c} n+l-1 \\ n-h \end{array} \right) (-1)^{h-n}
z^{-(n+l)} 
\sum_{m=1-h}^{l}
\left( \begin{array}{c} l+h-1 \\ m+h-1 \end{array} \right)
z^{l-m} (S_m \otimes \bbbone)\,. \nonumber
\end{eqnarray}
Using (\ref{zero}) we can rewrite the $(\bbbone\otimes S_0)$ term,
and find after a short calculation
\begin{equation}
\label{quotient}
(S_{-n} \otimes \bbbone) \cong
\left( \begin{array}{c} n-1 \\ n-h \end{array} \right) (-1)^{h-n}
\left\{ z^{h-n} (S_{-h} \otimes \bbbone) 
+ \sum_{m=1-h}^{-1} z^{-(m+n)} C_m (S_m \otimes \bbbone) \right\} \,,
\end{equation}
where
\begin{equation}
C_m = \sum_{l=m}^{-1}
\left( \begin{array}{c} l+h-1 \\ m+h-1 \end{array} \right) 
\left( 1 - \frac{(n+l-1)! \; (h-1)!}{(n-1)! \;(l+h-1)!} \right) \,.
\nonumber
\end{equation}

For completeness we should also give the result for $h=1$, where the
analysis simplifies to
\begin{equation}
\label{quotient1}
(T_{-n} \otimes \bbbone) \cong - (-z)^{-n} 
(\bbbone \otimes T_0) \cong (- z)^{-n+1} (T_{-1} \otimes \bbbone) \,.
\end{equation}

Taking $n=h+1$, we note that (\ref{quotient}) and (\ref{quotient1})
become 
\begin{equation}
0 \cong \sum_{m=1}^{h+1} 
\left( \begin{array}{c} h \\ h+1-m \end{array} \right)
z^{m-h-1} (S_{-m}\otimes\bbbone) \,, \nonumber
\end{equation}
which, for $z=1$, just reproduces the formula of Zhu \cite{Zhu}. 
Here we have used
\begin{equation}
\sum_{l=1}^{m} l 
\left( \begin{array}{c} h-1-l \\ h-1-m \end{array} \right)
= \left( \begin{array}{c} h \\ h+1-m \end{array} \right) \,,
\end{equation}
(see {\it e.g.}\ \cite[p. 176]{Formula}).

{}From our definition of $A(\H_0)$ it is clear that every highest weight
representation of $\H_0$ gives rise to a representation of $A(\H_0)$
with respect to the product structure induced by
$\Delta_{z,0}(S_0)$. As our space is at most as large as the space of
Zhu, it is then clear that our definition has to agree with the one of
Zhu. 
\medskip

The fermionic case is slightly simpler, as there are no zero modes,
and thus there is no relation corresponding to (\ref{zero}). We
therefore only have to calculate for $n\geq h$
\begin{eqnarray}
( S_{-n} \otimes \bbbone) & = & 
\widetilde{\Delta}_{0,-z} (S_{-n}) 
- \varepsilon_1 \sum_{l=1-h}^{\infty} 
\left( \begin{array}{c} n+l-1 \\ n-h \end{array} \right) 
(-1)^{l+h-1} (-z)^{-(n+l)} 
\left(\bbbone \otimes S_{l}\right) \nonumber \\
& \cong &
\varepsilon_1 (-1)^{h-n} \sum_{l=1-h}^{-\half} 
\left( \begin{array}{c} n+l-1 \\ n-h \end{array} \right) 
z^{-(n+l)} \left(\bbbone \otimes S_{l} \right) \,. \nonumber
\end{eqnarray}
For $l=1-h, \ldots , -\half$, we have furthermore
\begin{equation}
\varepsilon_1 \left( \bbbone \otimes S_l \right) =
\Delta_{z,0}(S_l) - 
\sum_{m=1-h}^{l} \left( \begin{array}{c} l+h-1 \\ m+h-1
\end{array} \right)
z^{l-m} \left(S_{m} \otimes \bbbone\right) \,. \nonumber
\end{equation}
Hence we find for $n\geq h$
\begin{equation}
\label{quotientf}
\left(S_{-n} \otimes \bbbone \right) \cong
(-1)^{h-n+1} \sum_{m=1-h}^{-\half} z^{-(n+m)} D_m 
\left(S_m \otimes \bbbone \right) \,,
\end{equation}
where
\begin{equation}
D_m = \sum_{l=m}^{-\half} 
\left( \begin{array}{c} l+h-1 \\ m+h-1 \end{array} \right)
\left( \begin{array}{c} n+l-1 \\ n-h \end{array} \right) \,.
\nonumber
\end{equation}

Using (\ref{binid}) this formula simplifies for $n=h$ to 
\begin{equation}
0 \cong \sum_{m=0}^{h-\half} 
\left( \begin{array}{c} h-\half \\ m \end{array} \right)
z^{-m} (S_{m-h}\otimes\bbbone) \,, \nonumber
\end{equation}
which, for $z=1$, just reproduces the formula 
of Kac and Wang \cite{KacWang}. 
By the same reasoning as before, it is then clear that our definition
agrees with the one of Kac and Wang.
\medskip

In the case of the $N=2$ algebra, the only fermionic fields are
$G^{\pm}$ of conformal weight $h=3/2$. Because of (\ref{quotientf}),
all negative modes of $G^{\pm}_{-m}$ with $m\geq 3/2$ can be
eliminated in the quotient. On the other hand, $G^{\pm}_{-m}$ vanishes
on the vacuum for $m=\half$, and thus all $G^{\pm}$ modes can be
removed. Furthermore, using (\ref{quotient}) and (\ref{quotient1}),
all (negative) modes of $L$ and $T$ can be eliminated, except for
$T_{-1}$ and $L_{-2}, L_{-1}$. On the other hand, $L_{-1}$ vanishes on
the vacuum, and thus can be removed by commuting it through to the
right. The space $A(\H_0)$ is therefore a certain quotient space of
the space generated by $L_{-2}$ and $T_{-1}$.

We can then equally well describe $A(\H_0)$ as a quotient space of the
space generated by $h=L_{-2}+L_{-1}$ and $q=T_{-1}$; this formulation
has the advantage that the two generators commute, and that they can
be directly identified with the eigenvalues of the highest weight with
respect to $L_0$ and $T_0$ (by (\ref{zero})), since
\begin{equation}
(\bbbone\otimes L_0) \cong (h\otimes\bbbone) \,, \hspace*{3cm}
(\bbbone\otimes T_0) \cong (q\otimes\bbbone) \,.
\nonumber
\end{equation}
Thus $A(\H_0)$ is a quotient space of the space of polynomials 
in $h$ and $q$. 

Generically, this space is infinite dimensional, and to obtain some
restrictions, we have to use singular vectors in the vacuum
representation. It is clear that all bosonic descendants of
singular vectors do not give new information for the quotient, as we can
always replace negative modes of $L$ and $T$ by $h, q, L_{-1}$ and some
non-negative modes. Apart from the $L_{-1}$ contributions, these give
only restriction which contain, as a factor, restrictions from the
original singular vector. The $L_{-1}$ contributions simply correspond to
an infinitesimal shift in the insertion point $z$, and thus do not
give new restrictions either. For fermionic descendants, a similar
argument implies that the only descendants of potential interest are
\begin{equation}
\label{descendents}
G^{\pm}_{-\half} \Nu\ \,, \hspace*{3cm}
G^{+}_{-\half} G^{-}_{-\half} \Nu \,, \nonumber
\end{equation}
where $\Nu$ is a singular vector. It is clear that all three are
trivial for the first generic singular vectors of the vacuum
representation, $G^{\pm}_{-\half}\Omega=0$, but in general, they need
not be trivial.  

It follows from the embedding structure for the cases corresponding to
Fig.\ 2.1 and Fig.\ 2.2 that all singular vectors are descendants of
the generic singular vectors of the vacuum representation. We can thus
conclude that $A(\H_0)$ is isomorphic to a polynomial ring in two
independent variables, and thus, in particular, infinite dimensional.
This shows that the corresponding theories are not rational.

In the case of the diagram of Fig.\ 2.3, there exists an additional
independent bosonic singular vector $\Nu$\footnote{To avoid confusion,
we should point out that $L_{-1}\Omega$ is a descendent of the two
generic singular vectors. We should also note that we implicitly
assume here, that the vacuum representation does not possess any
subsingular vectors which might give additional relations.}. The
relations of the generic singular vectors have already been taken into
account in the above derivation, and $A(\H_0)$ is therefore only
finite dimensional (and the corresponding theory rational) if $\Nu$
gives rise to {\it two independent} relations. For a bosonic singular
vector $\Nu$, only the third descendant in (\ref{descendents}) can
contribute, as the other two have odd fermion number and thus are
equivalent to zero in the quotient.

We conclude from this that the potential rational models all have a
non-trivial bosonic singular vector in the vacuum representation. Whether
the model is actually rational depends then on whether the 
$G^{+}_{-\half}G^{-}_{-\half}$ descendent gives an independent
relation or not. We have calculated the relations coming from $\Nu$ and
the $G^{+}_{-\half}G^{-}_{-\half}$ descendent for a few examples
explicitly. The first bosonic singular vector is given in each case as 
\begin{eqnarray}
&\Nu_{c=1} =& (2 L_{-2} - 3 T_{-1}T_{-1})\Omega \,, \nonumber \\
&\Nu_{c=\thalf}  = &
               (10 T_{-3} - 
                 3 L_{-3} + 
                 3 G^+_{-3/2} G^-_{-3/2}  - 
                12 L_{-2} T_{-1}  + 
                 8 T_{-1}T_{-1}T_{-1})\Omega \,, \nonumber\\
&\Nu_{c=-6} = &
            (-10 T_{-3}  - 
             6  L_{-3}  + 
             6 G^+_{-3/2} G^-_{-3/2} + 
             6 L_{-2}  T_{-1}  + 
             T_{-1}T_{-1}T_{-1})\Omega \,, \nonumber \\
&\Nu_{c=-1} =& 
           (42 T_{-4} +
            24  L_{-4} +
            27 T_{-2} T_{-2} -
            84 T_{-3} T_{-1} -
            6  G^+_{-3/2} G^-_{-5/2} \nonumber \\
          &&+ 6  G^+_{-5/2} G^-_{-3/2} -
            32 L_{-2} L_{-2} -
            36 L_{-3} T_{-1} +
            36 T_{-1} G^+_{-3/2} G^-_{-3/2}  \nonumber\\
          &&+ 12 L_{-2} T_{-1}  T_{-1} +
            9  T_{-1}  T_{-1}  T_{-1}  T_{-1} )\Omega \,,\nonumber\\
&\Nu_{c=-12} =&
          (-240 L_{-5}  
           +360 G^+_{-3/2} G^-_{-7/2}  
           +120 G^+_{-5/2} G^-_{-5/2}
           +840 L_{-2} T_{-3}  
           \nonumber\\
         &&+360 G^+_{-7/2} G^-_{-3/2} 
           +600 L_{-3} L_{-2}  
           +120 L_{-3} T_{-2}  
           +180 L_{-4} T_{-1}   
           \nonumber\\
         && -60 T_{-1} G^+_{-3/2} G^-_{-5/2}  
            +60 T_{-1} G^+_{-5/2} G^-_{-3/2}  
           -600 L_{-2} G^+_{-3/2} G^-_{-3/2}  
           \nonumber\\
         &&-300 L_{-2} L_{-2} T_{-1} 
            +60 L_{-3} T_{-1} T_{-1}  
            -60 T_{-4} T_{-1}
            +30 T_{-2} T_{-2} T_{-1}  
            \nonumber\\
         && -60 L_{-2} T_{-1} T_{-1} T_{-1} 
           +180 T_{-3} T_{-1} T_{-1} 
            -60 T_{-1} T_{-1} G^+_{-3/2} G^-_{-3/2}  
            \nonumber\\
         &&  -3 T_{-1} T_{-1} T_{-1} T_{-1} T_{-1}  
          -1992 T_{-5}
         )\Omega \,. \nonumber
\end{eqnarray}
The singular vector and its descendent give rise to the polynomial
relations (in $h$ and $q$) $p_1$, and $p_2$, respectively.
The algebra $A(\H_0)$ is then given by
$$ A(\H_0) = \C[h,q]/<p_1(h,q),p_2(h,q)>.$$
Our results for the five cases above are contained in Table 3.1.
$$
  \begin{array}
      {|c|c|c|c|}
       \hline
  c       & p_1(h,q)& p_2(h,q) \\
       \hline
  1       &(2 h - 3 q^2)    
          &(-2 h + q)(1 + 3 q)\\
       \hline
  3/2     &q (1 - 12 h + 8 q^2)
          &(2 h - q) (1 - 2 h + 5 q + 8 q^2) \\
       \hline
  -6      &q (2 + 6 h + q^2)
          &(2 h - q) (2 + 6 h + q^2) \\
       \hline
  -1      &(-4 h + 3 q^2) (1 + 8 h + 3 q^2)
          &(2 h - q) (2 + 3 q) (1 + 8 h + 3 q^2)\\
       \hline
  -12     &q (4 + 10 h + q^2) (6 + 10 h + q^2)
          & (-2 h + q) (4 + 10 h + q^2) (6 + 10 h + q^2)\\         
       \hline
   \end{array}
$$
\centerline{Table 3.1: Polynomials determining $A(\H_0)$ for 
                       certain values of $c$}

\vskip 0.3cm
We note that for the unitary models $c=1$ and $c=\frac32$, $A(\H_0)$ 
is finite-dimensional, as the two relations are independent. In the
other three cases, however, the two relations contain a common factor,
and thus $A(\H_0)$ is infinite dimensional.

\section{The coset argument}

We have shown in the last section that the $N=2$ super Virasoro
algebra is not rational for certain non-unitary cases.  In this
section we will analyse the remaining cases with $c<3$. The analysis
for $c\geq 3$, using the modular properties of the vacuum character,
is contained in the appendix~B.

We shall use the coset realisation (\ref{coset}) of the $N=2$ super
Virasoro algebra to show that certain admissible representations of
$\suh_k$ give rise to infinitely many irreducible representations in
the non-unitary cases.  The basic idea of this argument is due to Ahn
{\it et al.} \cite{Ahn}.  We shall present the argument in the more
general setting of the Kazama-Suzuki models as the counting argument
generalises. 
\smallskip

Recall that the Kazama-Suzuki models can be constructed from hermitian 
symmetric spaces \cite{KS}. More precisely, it has been shown in 
ref. \cite{KS} that if $G/H$ is a hermitian symmetric space, the coset
\begin{equation}
\label{KaZu}
\frac{\hat \frakg_k \oplus (\Fer)^{2n}}{\hat \frakh_k}\, ,
\end{equation}
where $n=\rank(G)=\rank(H)$,
contains the $N=2$ super Virasoro algebra and the explicit form 
of the $N=2$ super Virasoro generators $T,G^\pm,L$ in terms of the
$2n$ free  fermions and the $\hat \frakg_k$ currents has been given.
A complete list of hermitian symmetric space can, for example, be found 
in ref. \cite[Table 1]{KS}.
Note that for all hermitian symmetric spaces $\rank(G) = \rank(H)$,
$\frak g$ is simple and that $\frak h$ is of the form 
$\frak h = u(1) \oplus \frakh_1$, where $\frakh_1$ is semisimple. 
The case of the $N=2$ super Virasoro algebra corresponds to 
$\frakg = su(2)$ and $\frakh = u(1)$.

Before proceeding, we should note that it is in general rather
difficult to determine the actual coset algebra of a given coset. In
particular, even if the coset algebra is correctly identified for
generic level (which is usually a tractable problem, for example by
comparing generic characters), it is {\it a priori} not clear that
this identification remains correct at arbitrary level. However, for
the case of $\frakg = su(2), \frakh = u(1)$, the character
calculations of appendix~A show that the coset is indeed the $N=2$
super Virasoro algebra for arbitrary level $k$\footnote{This argument
relies on the conjectured embedding diagrams of the $N=2$ algebra.}. 
\medskip 

Using the explicit form of the $\uh$ current $T$ and the
Virasoro field $L$ in the coset \cite[eq.\ (4.5)]{KS} it is easy 
to obtain a formula for the eigenvalues $q$ and $h$ of 
$T_0$ and $L_0$, respectively,  acting on the subspace 
of the $\hat \frakg_k$ highest weight 
space with $\frak h$-weight $\lambda$ of a 
$\hat\frakg_k\oplus\Fer^{2n}$ highest weight 
representation
\begin{equation}
\label{qunu}
\begin{array}{lcl}
h  & = & {\displaystyle \frac{C_2(\frakg)}{2(k+g)} - 
   \frac{(\lambda,
          \lambda+2\rho_\frakh)}{2(k+g)} }
\vspace*{0.3cm}  \\
q & = & {\displaystyle -\frac{2}{k+g} 
           (\rho_\frakg-\rho_\frakh,
             \lambda) \,.}
\end{array}
\end{equation}
Here $C_2(\frakg)$ denotes the second order Casimir of the $\frakg$
representation on the $\hat \frakg_k$ highest weight space, $g$ the
dual Coxeter number of $\frakg$, and $\rho_\frakg$ and $\rho_\frakh$
are half the sum of the positive roots of $\frakg$ and $\frakh$,
respectively. For the case of $\frakg=su(2)$, the formulae become
\begin{equation}
\label{qunusu2}
h =  \frac{j(j+1)}{(k+2)} -    \frac{m^2}{(k+2)} \hspace*{3cm}  
q =  -\frac{2 m}{k+2}\,,
\end{equation}
where $j$ and $m$ label the spin and the magnetic quantum number of
the corresponding $su(2)$ representation. If for admissible
$k\not\in\N$ the admissible representations of $\suh_k$ are
MCFT representations, it follows directly from these formulae that
there are infinitely many highest weight states, as was already
observed by Ahn {\it et al.} \cite{Ahn}. (For more details see below.)
This implies then directly that the corresponding theory is not
rational.

It has now been shown that the admissible representations of
$\suh$ are indeed MCFT representations \cite[Corollary
2.11]{DLM2}. For general $\frakg$, the corresponding result
is not yet known, but we believe it to be true.

Assuming this for the general case, the argument can be generalised as
follows: we note that $\frakh$ has one simple root less than $\frakg$
which we denote by $\alpha$, and, that the Dynkin index of $\alpha$ is
$1$. For an admissible but non-integer level $k$ there always exists
an admissible representation of $\hat \frakg_k$ whose Dynkin label
corresponding to the fundamental weight dual to $\alpha$ is fractional
(\cite[Theorem 2.1 (c)]{KacWaki2} see also \cite[p.\ 236]{MW}). This
representation has in particular an infinite dimensional highest
weight space. Let $\Lambda$ be the $\frakg$-weight of the highest
weight vector $v_\Lambda$ of this representation, and denote by
$\Lambda_\frakh$ the $\frakh$-weight of $v_\Lambda$. Furthermore, let
$E_\alpha$ be the step operator corresponding to $\alpha$. Then the
vectors $E_{-\alpha}^n\,v_\Lambda$ are highest weight vectors of
$\frakh$, and their $\frakh$-weights are given as $\lambda_n =
\Lambda_\frakh + n\mu$, where $\mu$ is a non-zero $\frakh$-weight and
$n+1\in\N$. This implies that the expression
$(\lambda_n,\lambda_n+2\rho_\frakh)$ is unbounded for $n\to\infty$ and
hence, that there are infinitely many different values for the
conformal weight $h$ in (\ref{qunu}). Thus the coset (\ref{KaZu}) has
infinitely many inequivalent representations. 

To relate the arguments for $\frakg=su(2)$ to the
results of \S3, let us parametrise the admissible level as
$k=p/p'-2$, where $p,p'$ are coprime positive integers and
$p\ge2$. The admissible representations are given by the $\suh_k$
weights \cite[p.\ 4958]{KacWaki}
\begin{equation}
\label{admiswh}
\lambda_{n,l} = (k-n+l(k+2))\Lambda_0 + (n-l(k+2))\Lambda_1  \,,
\end{equation}
where $n$ and $l$ run through $n=0,\dots,p-2,\ l=0,\dots,p'-1$, and
$\Lambda_0,\Lambda_1 $ are the fundamental weights of $\suh$. We note
that the spin $j$ of the $su(2)$ representation on the highest weight
space of the $\suh_k$ representation corresponding to $\lambda_{n,l}$
is given by $j = \half (n-l(k+2))$. In particular, for $k\in\N$ which
corresponds to the unitary case, the spin $j$ is always half-integral.
If $k\not\in\N$, the admissible representations corresponding to the
weights $\lambda_{n,l}$ with $l\not=0$ have an infinite dimensional
highest weight space as $2j+1\not\in\N$. These representations give
rise to infinitely many MCFT representations of the coset algebra,
thus showing that it cannot be rational. Indeed, eq.\ (\ref{qunusu2})
implies that
\begin{equation}
 h + \frac{k+2}{4} q^2 = \frac{j(j+1)}{k+2} \,,
\end{equation}
where $j$ is the spin of the $su(2)$ representation. This equation
gives precisely the common factors in Table 3.1 for $c=-6,-1$ and
$-12$, {\it i.e.}\ $k+2=p/p' = \frac23,\frac32$ and $\frac{2}{10}$,
where the spin $j$ corresponds to the admissible representations of
$\suh_k$ with infinite dimensional highest weight space which are
given by weights $\lambda_{n,l}$ with $l\not=0$. The additional
discrete representations of $A(\H_0)$ correspond to the $\suh_k$
representations with the weights $\lambda_{n,0}$ ($n=0,\dots,p-2$):
for example in the second case $c=-1$, $(h,q)=(\frac13,-\frac23)$
comes from $n=1$ and $(h,q)=(0,0)$ from $n=0$.
\smallskip

We know that all irreducible representations satisfying the conditions
given by the polynomials in Table 3.1 are MCFT representations, so in
particular there exists a continuum of MCFT representations in the
non-unitary cases. The above argument, however, only shows that those
representations satisfying (\ref{qunusu2}) can be obtained from the coset
construction. Furthermore, it is clear that only countably many
representations of the coset MCFT can be constructed from the
admissible $\suh_k$ representations. It therefore seems that
the remaining representations cannot be constructed using the coset
realisation.
\smallskip

Finally, let us mention that our findings are in perfect agreement
with the results obtained in ref. \cite{BH}. The authors of {\it loc.\
cit.}\ have investigated the representation theory of several
exceptional $N=2$ super $\cal W$-algebras from a completely different
point of view.  The only rational models they found were unitary and
even contained in the unitary minimal series of the $N=2$ super
Virasoro algebra.

\section{Conclusion}

In this paper we have analysed systematically the question
whether the rational theories of the $N=2$ superconformal
algebra are always unitary. We have used three independent
arguments to exclude the existence of rational non-unitary theories.
Where possible we have checked that the different methods
lead to consistent conclusions.

One of the methods is based on the coset realisation of the $N=2$
algebra, and we have already indicated in section~4 how this argument
can be generalised to the Kazama-Suzuki models. Apart from the
(aforementioned) problem that the admissible representations are not
yet known to be MCFT representations in general, this argument does
not exclude that the theories corresponding to non-admissible affine
theories are rational. However, we expect that there should be
`fewer' singular vectors than in the admissible case, and thus that
the corresponding theories should also not be rational. It should be
possible to settle both of these problems as soon as the
representation theory of the Kac-Moody algebras at non-integer level
is understood in detail. 
\smallskip

The coset argument is even more general, as it does not involve the
fermions. Indeed, ignoring the fermions where applicable, it can be
applied to cosets of the form
\begin{equation}
\label{gencoset}
\frac{\hat\frakg^{(1)}_{k_1}\oplus\dots
		\oplus\hat\frakg^{(n)}_{k_n}}
  {\hat\frakh^{(1)}_{l_1}\oplus\dots
		\oplus\hat\frakh^{(m)}_{l_m}} \,, 
\end{equation}
where the $\hat\frakg^{(i)}_{k_i}$ and $\hat\frakh^{(j)}_{l_j}$ are
simple affine Kac-Moody algebras at admissible level, and the sum of
the numbers of simple roots of $\frakg^{(i)}_{k_i}$ with
$k_i\not\in\N$, is bigger than the corresponding sum
for the denominator. For the arguments to work for general
(admissible) $k_i$, we also have to assume that one of the
$\hat\frakg^{(i)}$ with $k_i\not\in\N$ contains a
simple root of Dynkin index $1$ which is not contained in the
denominator.

In particular, the arguments apply to the diagonal coset
$$ \frac{\hat\frakg_{k_1}\oplus\hat\frakg_{k_2} }
        {\hat\frakg_{k_1+k_2}} \,,$$
indicating that they only give rise to rational models if at least one
of the levels $k_1$ or $k_2$ is a positive integer ({\it c.f.}\ the
conjecture in ref. \cite[p.\ 2421]{BEHHH}). This is for example the
case for the coset realisation of the (non-unitary) minimal models of
the Virasoro algebra where one has $\frakg = su(2)$ and $k_2=1$.
There are certain purely bosonic coset MCFTs which are of the general
form (\ref{gencoset}), {\it e.g.}\ those corresponding to the unifying
$\cal W$-algebras associated to the unitary series of the Casimir
$\cal W$-algebras ${\cal WA}_n$ or ${\cal WD}_n$ \cite[Table
7]{BEHHH}. Our arguments confirm in these cases the conjecture of
ref. \cite[p.\ 2422]{BEHHH} that all rational models of the unifying
$\cal W$-algebras are also rational models of the Casimir $\cal
W$-algebras (which are, in these cases, all unitary).
\medskip

Let us close by mentioning some open problems.  It would be
interesting to know under which conditions all representations of a
coset MCFT $\hat\frakg/\hat\frakh$ can be obtained from MCFT
representations of $\hat\frakg$ --- as we have seen in section~4, this
is not the case for certain non-rational $N=2$ models, where there
exists a continuum of representations.  In the same spirit it would be
interesting to describe all representations of the $N=2$ super
Virasoro algebra that can be obtained from the admissible $\suh_k$
representations for given $k$ and to investigate whether they define
quasi-rational theories in the sense of \cite{Nahm}. It would also be
important to have a more general criterion for determining whether a
coset MFCT is rational or not. Finally, in order to complete the
arguments for the general case, it would be necessary to have a
better understanding of the representation theory of affine Kac-Moody
algebras, in particular at admissible level.

\bigskip
\leftline{\bf Acknowledgements}
We would like to thank M.\ D\"orrzapf, H.\ Kausch, P.\ Goddard, 
A.\ Kent, J.\ Nekovar and  G.\ Watts for discussions, 
and A.\ Honecker and M.\ R\"osgen for comments on a draft version 
of this paper. We are grateful to G.\ Watts for pointing out an error
in an earlier version. We also thank the referee for pointing out the
relevance of ref. \cite{Ahn}.

W.\ E. is supported by the EPSRC, and M.\ R.\ G. is supported by a
Research Fellowship of Jesus College, Cambridge. We also acknowledge
partial support from PPARC and EPSRC, grant GR/J73322.


\appendix


\section{Calculation of vacuum characters}

In this appendix we want to calculate the vacuum characters of the
$N=2$ super Virasoro algebra from the embedding diagrams of \S2 
and from the coset realisation described in \S4.

Let us first calculate the vacuum characters of the $N=2$ super 
Virasoro algebra in the cases corresponding to the embedding diagrams 
in Fig. 2.1-2.3. In order to be able to determine the characters from
the embedding diagrams we have to assume that there do not exist
subsingular  vectors in the vacuum representation, {\it i.e.}\ vectors
which are not singular in the vacuum Verma module but become singular in 
the quotient of the vacuum Verma module by its maximal proper
submodule\footnote{In \cite{Gato} certain representations of the $N=2$
superconformal algebra have been found which possess subsingular
vectors. However, these representations are rather special and do not
include the vacuum representation.}. We also want to assume that the
character of a submodule of a 
Verma module generated from a level $n$ charged singular vector is 
given by $q^n/(1+q^{n-n'})$ where $n'<n$ is the level of the 
uncharged singular vector (or highest weight vector) connected by 
a line to the charged singular vector in the embedding diagram. 
This means for example that the character of the submodule 
generated from $G^+_{-\half}\Omega$ or $G^-_{-\half}\Omega$ is
just given by $q^{\half}/(1+q^{\half})$, which is obvious in this case. 

In the case of the embedding diagram shown in Fig.\ 2.1 and 
Fig.\ 2.2 all singular vectors are embedded in the two submodules 
generated from
$G^\pm_{-\half}\Omega$. Moreover, the embedding diagram implies that 
the intersection of the two submodules generated by
$G^+_{-\half}\Omega$ and $G^-_{-\half}\Omega$ is trivial.
Therefore, the vacuum character of the $N=2$ super Virasoro algebra
with $c(1,p')=3(1-2p')$ ($p'\not\in\Q$ or $p'\in\N$) is given by
\begin{equation}
\label{genchar}
\chi_{1,p'}(q) = 
q^{-c(1,p')/24} \prod_{n=1}^\infty \frac{(1+q^{n+\half})^2}{(1-q^n)^2}
\left( 1 - 2 \frac{q^{\half}}{1+q^{\half}} \right). 
\end{equation}

The case corresponding to the embedding diagram in Fig.\ 2.3 is more 
interesting. Here we have to subtract and add successively the 
characters of the modules generated by the corresponding singular
vectors. Using the two assumptions above we obtain that 
the vacuum character of the $N=2$ super Virasoro algebra
with $c(p,p')= 3(1-2\frac{p'}{p}); p,p'\in\N; (p,p')=1; p\ge2$
is given by 
\begin{eqnarray}
\chi_{p,p'}(q) &=& 
q^{-c(p,p')/24} \prod_{n=1}^\infty \frac{(1+q^{n+\half})^2}{(1-q^n)^2}
 \times \nonumber \\
&&\qquad\qquad\Big( 1 -
 \sum_{n=0}^\infty  
               q^{p'(n+1)(p(n+1)-1)}+
        2\frac{q^{pn(p'n+1)+p'n+\half}}{1+q^{pn+\half}} 
 \nonumber \\
&&\qquad\qquad\quad\
 +\sum_{n=1}^\infty  
               q^{p'n(pn+1)}+        
        2\frac{q^{pn(p'n-1)-p'n-\half}}{1+q^{pn-\half}} 
\Big). \nonumber
\end{eqnarray}

In particular, for $p'=1$ the above formula gives the vacuum character 
of the unitary minimal model with central charge $c= 3(1-\frac{2}{p})$
\footnote{Although the multiplicities in the embedding diagrams of
ref. \cite{Do,Ma,Ki} are not correct, the authors of {\it loc.cit.}\
obtained the correct characters in the unitary case.}.

Finally, note that the last expression for the vacuum character 
$\chi_{p,p'}$ can be rewritten as
\begin{equation}
\label{nuchar}
\chi_{p,p'}(q) = 
 q^{-c(p,p')/24} 
 \left(\prod_{n=1}^\infty \frac{(1+q^{n-\half})^2}{(1-q^n)^2} \right)
 \sum_{n\in \Z} q^{p'n(pn+1)} \frac{1-q^{pn+\half}}{1+q^{pn+\half}}.
\end{equation}
\medskip

In the second part of this appendix we use the coset realisation of 
the $N=2$ super Virasoro algebra (\ref{coset}) for the calculation of
the vacuum characters $\chi_{p,p'}$. Recall that the central charge of
the $N=2$ algebra is given as $c= \frac{3k}{k+2}$ ($k\not=0,-2$), and
that the vacuum representation is given by the space of all uncharged
$\uh$ highest weight states in the vacuum representation of
$\suh_k\oplus\Fer^2$. 

The character $\chi_{p,p'}$ is therefore the $\uh$ uncharged part of 
$\chi^{\suh}_k \chi^{\Fer^2}$ divided by the $\uh$ character 
\begin{equation}
\chi_{p,p'}(q) =  \frac{1}{\chi^{\uh}(q)}
 \Res_z\left( \frac{1}{z} \chi^{\suh}_k(q,z) \chi^{\Fer^2}(q,z)
       \right)\,, 
\label{Resn2}
\end{equation}
where  
$\chi^{\uh}(q) = 1/\eta(q) = q^{-\frac{1}{24}}/ \tilde\eta(q)$ and
$\eta(q) = e^{\frac{\pi i\tau}{12}} \tilde\eta(q) = 
             e^{\frac{\pi i\tau}{12}} \prod_{n=1}^{\infty} (1 - q^n)$.
(Here we have used the $\suh$ characters 
$\chi^{\suh}_k(q,z)=\mbox{tr} q^{L_0} z^{2J^3_0}$ 
which also take the zero mode of $J^3$ into account.) 

There exist two types of embedding diagrams for $\suh_k$ vacuum 
representations with level $k>-2$ (corresponding to $c<3$)
\cite[Lemma 4.1]{Mali}:
either the level $k$ can be written as $k=p/p'-2$, where $p,p'\in\N$,
$(p,p')=1$ and $p\ge2$, or the vacuum representation is generic, 
{\it i.e.}\ all singular vectors are descendents of the level zero singular
vector. In the former case the representation is admissible
and the vacuum character is given by \cite{KacWaki}
\begin{equation}
\label{adchar}
\chi^{\suh}_k(q,z) = 
   \frac{\vartheta_{p',pp'}(\tau,z/p') - \vartheta_{-p',pp'}(\tau,z/p')}
        {\vartheta_{1,2}(\tau,z) - \vartheta_{-1,2}(\tau,z)} \,,
\end{equation}
where
$$ \vartheta_{k,\lambda}(\tau,z) := 
   \sum_{n\in \Z+\frac{\lambda}{2k}} q^{kn^2} z^{2 k n}\,.
$$
(Note that our $z$ corresponds to $q^{z/2}$ in \cite{KacWaki}.) 
In the latter case the vacuum character is generic, and is 
given by
$$ \chi^{\suh}_k(q,z) = 
\frac{q^{-\frac{c}{24}}}
{\prod_{n=1}^{\infty} (1-q^n) (1 - q^n z^2) (1 - q^n z^{-2})} \,.$$

Let us first consider the generic ({\it i.e.}\ not admissible) case
with $c<3$. 
To evaluate the above residue (\ref{Resn2}) we want to use the 
following expression for the fermionic character 
\begin{equation}
\chi^{\Fer^2}(q,z)  =  q^{-\frac{1}{24}}
\prod_{n\geq 1} \left( 1 + q^{n - \half} z^2 \right) 
\left( 1 + q^{n - \half} z^{-2} \right) 
 =  \frac{1}{\eta(q)} 
\sum_{m\in\Z} q^{\half m^2} z^{2m}   \nonumber
\end{equation}
which follows from the product formula for the
$\vartheta_3$ function (see {\it e.g.}\ \cite[p.\ 164]{Hi}) 
\begin{equation}
 \vartheta_3(\tau/2,z) = \sum_{n\in\Z} q^{\frac12 n^2} z^{2n}= 
     \prod_{n=1}^\infty (1-q^{n}) (1-z^2 q^{n-1/2})(1-z^{-2}
q^{n-1/2}).
\label{theta3}
\end{equation}

Furthermore, we shall also use the following identity for the 
denominator of the $\suh$ character (see {\it e.g.}\ 
\cite[p.\ 262 eq.\ (5.26)]{KP}) 
\begin{equation}
\frac{1}{\prod_{n=1}^{\infty} (1 - q^n z^2) (1 - q^n z^{-2})}  = 
\frac{1}{\tilde\eta(q)^2} \sum_{l\in\Z} \phi_l(q) z^{2l} \,, 
\nonumber
\end{equation}
where
$$ \phi_l(q) = \sum_{r=0}^{\infty} (-1)^r q^{lr + \half r (r+1)} \,.
$$
An important property of $\phi_l(q)$ is that
$\phi_{-l}(q) = q^l \phi_l(q).$

The vacuum character of the $N=2$ model is then (up to 
the $q^{-c(1,p')/24}$ term)
$$ - \Res_z\left\{ \frac{1}{\tilde\eta(q)^3} \sum_{m,l\in\Z} 
q^{\half m^2} \phi_l(q) 
\left( z^{2l} z^{2m} ( z^{1} - z^{-1}) \right)
\right\} \,,$$
where $p'\in\N$ or $p'\not\in\Q$. The evaluation of the residue gives
\begin{eqnarray}
\frac{1}{\tilde\eta(q)^3} \sum_{l\in\Z} \left( q^{\half l^2}
\phi_l(q) - q^{\half (l+1)^2} \phi_l(q) \right) 
& = & \frac{1}{\tilde\eta(q)^3} \sum_{l\in\Z} q^{\half l^2} 
\left( \phi_l(q) - q^{-l+\half} \phi_{-l}(q) \right) \nonumber \\
& = & \frac{1}{\tilde\eta(q)^3} \sum_{l\in\Z} q^{\half l^2} 
\phi_l(q) (1-q^{\half}) \nonumber \\
& = & \frac{1}{\tilde\eta(q)^3} (1-q^{\half}) \sum_{l\in\Z} 
\sum_{r=0}^{\infty} (-1)^r q^{\half (r^2 + r + 2 l r + l^2)} 
\nonumber \\
& = & \frac{1}{\tilde\eta(q)^3} (1-q^{\half}) \sum_{\hat{l}\in\Z} 
\sum_{r=0}^{\infty} (-1)^r q^{\frac{r}{2}} q^{\half \hat{l}^2} \,, 
\nonumber 
\end{eqnarray}
where $\hat{l}=l+r$. We can now do the sums over $\hat{l}$
and $r$, and obtain the $N=2$ vacuum character 
$$ \chi_{1,p'}(q) = q^{-\frac{c(1,p')}{24}} 
\left(\prod_{n=1}^\infty \frac{(1+q^{n-1/2})^2}{(1-q^n)^2} \right) 
\left( 1 - 2 \frac{q^{\half}}{1+q^{\half}} \right)\,, $$
where we have used that
$
\frac{1 - q^{\half}}{1+q^{\half}} = 1 - 2
\frac{q^{\half}}{1+q^{\half}}.$
This is indeed the generic $N=2$ vacuum character, where the only null
vectors are $G^{\pm}_{-\half} \Omega$ ({\it c.f.}\ eq.\
(\ref{genchar})).

Finally, consider the case where the $\suh_k$ vacuum character
is admissible, {\it i.e.}\ $k=p/p'-2$ with $p,p'\in\N$, $(p,p')=1$ 
and $p\ge2$. In this case we find, using (\ref{adchar}) and the 
well-known denominator formula,
$$ \frac{\chi^{\suh}_k(q,z)}{\chi^{\uh}(q)} = 
\frac{q^{-\frac{c(p,p')-1}{24}}}
{\prod_{n=1}^{\infty} (1 - q^n z^2) (1 - q^n z^{-2})}
\sum_{n\in\Z} q^{n p'(1+ pn)} 
\frac{( z^{2pn+1} - z^{-2pn -1})}{z^2 -1} 
\,.$$

Using (\ref{Resn2}), the $N=2$ vacuum character is then
(up to the $q^{-c(p,p')/24}$ term which we suppress for the moment) 
the residue   
$$ - \Res_z\left\{ \frac{1}{\tilde\eta(q)^3} \sum_{n,m,l\in\Z} 
q^{n p' (1 + pn)} q^{\half m^2} \phi_l(q) 
\ z^{2l} z^{2m} ( z^{2pn + 1} - z^{-2pn-1}) \
\right\} = (*)  \,.$$
The $z$-dependent part is
$ \Res_z\left( z^{-1 + 2 (l+m-pn)} -z^{1+2(pn + m + l)} \right),$
whose residue is easily obtained. By expressing $m$ in terms of
$n$ and $l$, the sum then becomes
$$(*) = \frac{1}{\tilde\eta(q)^3} \sum_{n,l\in\Z}
\left\{  q^{ n p'(1+pn)} \phi_l(q) q^{\half (l-pn)^2}
-  q^{ n p'(1+pn)} \phi_l(q) q^{\half (l+pn+1)^2} \right\}
\hspace*{1.0cm} 
$$
\begin{eqnarray}
& = & \frac{1}{\tilde\eta(q)^3} \sum_{n,l\in\Z}
\left[ q^{ n p'(1+pn)} q^{\half(l^2-2lpn+p^2 n^2)} \phi_l(q) \right.
\nonumber \\
& & \hspace*{4cm} \left.
- q^{ n p'(1+pn)} q^{\half(l^2+p^2 n^2+1+2l+2pn+2pnl)}
\phi_l(q)\right] \nonumber \\
& = & \frac{1}{\tilde\eta(q)^3} \sum_{n,l\in\Z}
\left[ q^{ n p'(1+pn)} q^{\half(l^2+p^2 n^2 - 2 lpn)} 
\Bigl(\phi_l(q) - q^{\half(1-2l + 2pn)} \phi_{-l}(q)\Bigr) 
\right] \nonumber \\
& = & 
\frac{1}{\tilde\eta(q)^3} \sum_{n,l\in\Z}
\left[ q^{ n p'(1+pn)} q^{\half(l^2+p^2 n^2 - 2 lpn)} 
\phi_l(q) \Bigl(1 - q^{pn + \half}\Bigr) \right] \,, \nonumber
\end{eqnarray}
where we have replaced $l$ by $-l$ in the second sum of the
penultimate line, and used the previously mentioned symmetry of
$\phi_l$ in the last equation. Next we use the explicit expression for
$\phi_l$ to obtain
$$ (*) = 
\frac{1}{\tilde\eta(q)^3} \sum_{n\in\Z}
q^{ n p'(1+pn)} \Bigl(1 - q^{pn + \half}\Bigr) \sum_{l\in\Z} 
\sum_{r=0}^{\infty} (-1)^r q^{\half(r^2 + r + l^2 + p^2 n^2 + 2 lr 
- 2 l p n )}\,.
$$
The last exponent of $q$ can be rewritten as
$$
\half\Bigl(r^2 + r + l^2 + p^2 n^2 + 2 lr - 2 l p n \Bigr) = 
\half\Bigl(\hat{l}^2 + r + 2 r p n\Bigr) \,,
$$
where $\hat{l} = l - p n + r$. We then replace the sum over $l$ by a
sum over $\hat{l}$ which gives
$\sum_{\hat{l}\in\Z} q^{\half \hat{l}^2} = 
\tilde\eta(q) \prod_{m=1}^{\infty} (1 + q^{m - \half})^2$.
The sum over $r$ is the geometric series 
$\sum_{r=0}^{\infty} (-1)^r q^{\half(r(1+2pn))} 
  = 1/(1 + q^{pn + \half})$, and we thus arrive at (compare for
example \cite{RY,HNY,Ahn}) 
$$ \chi_{p,p'}(q) = q^{\frac{-c(p,p')}{24}} 
\left(\prod_{m=1}^\infty \frac{(1+q^{m-\half})^2}{(1-q^n)^2} \right)
\sum_{n\in \Z} q^{pp'n^2+p'n}
\frac{1-q^{pn+\half}}{1+q^{pn+\half}}\,.$$
This expression equals the one derived from the embedding diagram
(\ref{nuchar}). 


\section{Modular properties of the vacuum characters}

As already mentioned in \S2, it was shown in \cite{Zhu} that for
bosonic rational conformal field theories the space of torus
amplitudes which is invariant under the natural action of the modular
group is finite dimensional. 

We expect therefore that the dimension of the space spanned by the
functions $\chi\vert_A(\tau)=\chi(A\tau)$ ($A\in\SL$), where $\chi$ is
the vacuum character, is finite if and only if the $N=2$ super
Virasoro algebra is rational for the corresponding value of $c$. We
shall show, using the following two lemmas, that the dimension of this
space is infinite for $c\ge3$ and for the non-unitary models
corresponding to the embedding diagrams in Fig.\ 2.1 and Fig.\ 2.2,
{\it i.e.}  $c = c(1,p')$ with $p=1,p'\not\in\Q$ or $p=1,p'\in\N$. On
the other hand, it is finite for the unitary models with $c = c(p,1)$
and $p'=1,p\ge3$. (For $p'=1,p=2$ the dimension is clearly 1 since
$\chi_{1,2} = 1$.)

\begin{lma}
For $k=p-2\in\N$ the vacuum characters $\chi_k(\tau) := \chi_{p,1}(q)$
($ q = e^{2\pi i \tau}$) of the 
$N=2$ super Virasoro algebra are modular functions on $\Gamma(24k(k+2))$. 
More explicitly, they are given by 
$$\chi_k(\tau) = \frac{1}{\eta^3(\tau)}
                 \sum_{m\bmod 2k}
                    \Theta_{L,(\frac{1}{2(k+2)},\frac{m}{2k})}(\tau)
                    \vartheta_{m(k+2),k(k+2)}(\tau/2) \,,
$$
where the 
$\vartheta_{\lambda,k}= \sum_{n\in\Z} q^{\frac{(2kn+\lambda)^2}{4k}}$
are Riemann-Jacobi theta functions 
and the $\Theta_{L,\mu}$ are Hecke indefinite modular forms (of weight one)
associated to the lattice $L = \Z\oplus\Z$ and the quadratic form 
$Q(\gamma) = 2(k+2)\gamma_1^2 - 2k\gamma_2^2$.
\end{lma}
\bigskip

\noindent {\it Proof.} We first recall the definition of a Hecke
indefinite modular form 
(see \cite{He} or \cite[pp.\ 254]{KP} for more details). 
Let $L\subset\R^2$ be a lattice of rank two and $Q:L\to 2\Z$
an indefinite quadratic form such that $Q(x) = 0, x\in L$ implies
$x=0$.  Denote by $L^\sharp$ the lattice dual to $L$, 
$L^\sharp = \{ x\in \R^2 \vert B(x,y) \in \Z \ \mbox{for}\ y\in L\}$,
where $B(\gamma,\gamma')=\frac12(Q(\gamma+\gamma')-Q(\gamma)-Q(\gamma'))$ 
is the bilinear form associated to $Q$.
Let $G_0$ be the subgroup of the identity component 
of the orthogonal group of $(B,\R^2)$ which preserves $L$ and fixes all 
elements of $L^\sharp/L$. Fix a factorisation 
$Q(\gamma) = l_1(\gamma) l_2(\gamma)$ where $l_1$ and $l_2$ are real
linear and set $\sign(\gamma) = \sign(l_1(\gamma))$.
Then 
$$\Theta_{L,\mu}(\tau) := 
   \sum_{ {{\gamma\in L+\mu} \atop {B(\gamma,\gamma)>0}}
          \atop {\gamma\ \mbox{mod}\ G_0} }
    \sign(\gamma) q^{Q(\gamma)/2}
$$
is called a Hecke indefinite modular form associated to $\mu$ and $L$.
It is a modular form of weight one on $\Gamma(N)$, where
$N\in\N$ satisfies $N Q(\gamma)\in 2\Z$ for all $\gamma\in L^\sharp$. 

The case we are interested in has been studied in \cite[pp.\ 256]{KP}.
We have $L = \Z\oplus\Z$ and 
$$Q(\gamma) = 2(k+2)\gamma_1^2 - 2k\gamma_2^2
  = l_1(\gamma) l_2(\gamma) \,, $$
where $l_1(\gamma) = \sqrt{2(k+2)}\gamma_1+\sqrt{2k}\gamma_2$ 
and   $l_2(\gamma) = \sqrt{2(k+2)}\gamma_1+\sqrt{2k}\gamma_2$
so that $Q(\gamma)=0$ for $\gamma\in L$ implies $\gamma = 0$. 
Then $B$ is given by 
$$ B(\gamma,\gamma') = 2(k+2)\gamma_1\gamma_1' -2k\gamma_2\gamma_2'\,,$$
implying that $L^\sharp$ equals $\frac{1}{2(k+2)}\Z\oplus \frac{1}{2k}\Z$.
We observe that $A$, given by 
$$A (\gamma_1,\gamma_2) = ( (k+1)\gamma_1 +     k\gamma_2,
                            (k+2)\gamma_1 + (k+1)\gamma_2 ) \,,$$
satisfies $Q(\gamma) = Q(A\gamma)$, and that the group generated by $A$
is the identity component of the orthogonal group of $(B,\R^2)$ which
leaves $L$ invariant. Furthermore, $A^2$ generates $G_0$.
Hence the functions $\Theta_{L,(\frac{1}{2(k+2)},\frac{m}{2k})}$ 
($m\in \Z$) are modular forms of weight one on $\Gamma(4k(k+2))$.

Moreover, one can show that the Hecke indefinite modular forms 
$\Theta_{L,(\frac{1}{2(k+2)},\frac{m}{2k})}$ are given by
\cite[p.\ 258]{KP}
\begin{eqnarray}
  \Theta_{L,(\frac{1}{2(k+2)},\frac{m}{2k})}(\tau) & =  
    & \left( \sum_{s\ge,n\ge0}  -\sum_{s<0,n<0} \right)
        (-1)^s q^{(k+2)(\frac{s}{2}+n+\frac{1}{2(k+2)})^2-
              k  (\frac{s}{2}+\frac{m}{k})^2}          \nonumber \\
  & - &\left( \sum_{s\ge0, n>0}  -\sum_{s<0,n\le0} \right)
       (-1)^s q^{(k+2)(\frac{s}{2}+n-\frac{1}{2(k+2)})^2-
              k  (\frac{s}{2}+\frac{m}{k})^2} \,.          \nonumber
\end{eqnarray}
We thus find
\begin{equation}
\begin{array}{l}
{\displaystyle \frac{1}{\eta^3} \sum_{m\bmod 2k} 
      \Theta_{L,(\frac{1}{2(k+2)},\frac{m}{2k})}(\tau)
       \vartheta_{m(k+2),k(k+2)}(\tau/2) }              \\
{\displaystyle \qquad = \frac{1}{\eta^3}
      \sum_{m'\in \Z} \left( \sum_{s\ge,n\ge0}  -\sum_{s<0,n<0}   \right)
        (-1)^s q^{(k+2)(\frac{s}{2}+n+\frac{1}{2(k+2)})^2-
                    k  (\frac{s}{2}+\frac{m'}{k})^2+
                   \frac{k+2}{2k}{m'}^2} } \\
{\displaystyle \qquad -\frac{1}{\eta^3}
     \sum_{m'\in \Z} \left( \sum_{s\ge0, n>0}  -\sum_{s<0,n\le0} \right)
       (-1)^s q^{(k+2)(\frac{s}{2}+n-\frac{1}{2(k+2)})^2-
                    k  (\frac{s}{2}+\frac{m'}{k})^2+
                   \frac{k+2}{2k}{m'}^2}}        \\
{\displaystyle \qquad = q^{-\frac{3}{24}+\frac{1}{4(k+2)}}
     \prod_{n=1}^\infty \frac{(1+q^{n+1/2})^2}{(1-q^n)^2} \times}
                                                     \\
{\displaystyle \qquad\qquad\Big(\quad 
        \left( \sum_{s\ge,n\ge0}  -\sum_{s<0,n<0}   \right)
           (-1)^s  q^{s((k+2)n+\frac12)+(k+2)n^2+n} } \\
{\displaystyle \qquad\qquad  
      \  -\left( \sum_{s\ge0, n>0}  -\sum_{s<0,n\le0} \right)
           (-1)^s  q^{s((k+2)n-\frac12)+(k+2)n^2-n}
     \quad \Big)}                                         \\
{\displaystyle \qquad = q^{-\frac{c(k+2,1)}{24}}
     \prod_{n=1}^\infty \frac{(1+q^{n+1/2})^2}{(1-q^n)^2}
     \sum_{n\in \Z} 
           q^{(k+2)n^2+n} 
           \frac{1-q^{(k+2)n+1/2}}{1+q^{(k+2)n+1/2}}\,,}
\end{array} 
\end{equation}
where we have used the definition of the Riemann-Jacobi theta
functions $\vartheta_{m,k} = \vartheta_{m+2k\Z,k}$ and 
$\Theta_{L,\mu}=\Theta_{L,\mu+(0,\Z)}$ in the second equality, and 
the well-known product formula for the $\vartheta_3$ function
(\ref{theta3}) for $\tau/2$ and $z=1$ in the third. 
The last expression equals (\ref{nuchar}) for $p=k+2,p'=1$.

The property that the $\chi_k$ are modular functions on 
$\Gamma(24k(k+2))$ follows now directly from the well-known 
modular properties of the Riemann-Jacobi theta functions and the 
$\eta$ function.
\qed
\bigskip

It is then clear that the space spanned by $\chi_{p,1}\vert_A$
($A\in\SL$) is finite dimensional, as $\Gamma(24p(p-2))$ has  
finite index in $\SL$ for $p\ge 3$.
\smallskip

In order to prove that the space spanned by the functions
$\chi_{1,p'}(A\tau)$ ($A\in \SL$) is infinite dimensional, we need the
following lemma:

\begin{lma} 
Let $f:\C\to\C$ be a function of the form 
$f(\tau) =q^\alpha  P(q)/Q(q)$, where $\alpha\in\Q$,
$P$ and $Q$ polynomials and $q=e^{2\pi i\tau}$. 
Then for any $N>0$ and $k\in \Z$, the space spanned by 
the functions $f\vert_{k,A}$ ($A\in\Gamma(N)$) is 
infinite dimensional if $f$ is not constant. 
Here $f\vert_{k,A}$ is defined as
$$f\vert_{k,A}(\tau) = (c\tau+d)^{-k} f(A\tau)\,,$$
where $A=\left(\matrix{a & b \cr c & d} \right)$ and 
$A\tau = \frac{a\tau+b}{c\tau+d}$.
\end{lma}
\bigskip

\noindent {\it Proof.} Assume that the space spanned by the
$f\vert_{k,A}$ ($A\in\SL$) is $n$ dimensional, where $n<\infty$, and
that $f$ is not constant. Let $\tilde\tau$ be $\tau/s$ where $s$ is
the denominator of $\alpha=\frac{r}{s}$, $\tilde q = e^{2\pi i \tilde
\tau}$, and let $\tilde P$, $\tilde Q$ be the polynomials given by 
\begin{eqnarray}
    \tilde P (\tilde q) &= q^\alpha P(q),\quad 
    \tilde Q (\tilde q) &=\quad\ \  Q(q)
    \qquad\mbox{for}\ \alpha\ge0  \nonumber\\
    \tilde P (\tilde q) &= \quad P(q),\quad  
    \tilde Q (\tilde q) &= q^{-\alpha} Q(q)
    \qquad\mbox{for}\ \alpha<0\,.
   \nonumber 
  \end{eqnarray}
Then there exist $n$ matrices $A_i$ ($i=1,\dots,n$)
such that the functions $f\vert_{k,A_i}$ ($i=1,\dots,n$) are linearly
dependent over $\C$. (Without loss of generality we can assume that
$A_i^{-1} A_j$ are not of the form $\left(\matrix{* & * \cr 0 & *}
\right)$ as we are interested in a basis over $\C$). Hence the
polynomials $\tilde P(\tilde q_j) \prod_{i\not=j} \tilde Q(\tilde
q_i)$  ($j=1,\dots,n$) with $\tilde q_i = e^{2\pi i A_i\tilde\tau}$
are  linearly dependent over $\C[\tilde\tau]$, and thus the
$\tilde q_j$ are algebraically dependent over $\C[\tilde\tau]$.
Applying $A_1^{-1}$ to $\tilde\tau$ we can assume that $A_1$ is the
identity.  Looking at the asymptotic behaviour of the $\tilde q_i$ for
$\tau\to -i\infty$ we observe that there cannot be a term containing
$\tilde q_1$.  By induction on $n$ we find that the $\tilde q_i$ are
algebraically  independent. This gives the desired contradiction.  
\qed
\bigskip

The proof of lemma B.2 is due to J. Nekovar \cite{Nekovar}. 
\smallskip

The last lemma proves that the space spanned by the functions
$\chi_{1,p'}(A\tau)$ ($A\in \Gamma(48)$) is infinite dimensional since
the function $\frac{\eta((\tau+1)/2)^2}{\eta(\tau)^4}\,
\chi_{1,p'}(\tau)$ satisfies the assumptions of the lemma and
$\frac{\eta((\tau+1)/2)^2}{\eta(\tau)^4}$ is invariant under the
$\vert_{-1,A}$ action for $A\in\Gamma(48)$.  Therefore, the space
spanned by the functions $\chi_{1,p'}(A\tau)$ ($A\in \SL$) is infinite
dimensional.  We also expect that the dimension of the corresponding
space is infinite for $c = c(p,p')$ with coprime integers $p',p \ge
2$.

Finally the embedding diagrams of the vacuum Verma modules for $c\ge3$
({\it c.f.}\ the end of \S2) imply that the corresponding vacuum
characters are given by the product of the generic Verma module
character and a rational function of $q^\half$. We can therefore again
apply lemma B.2 to conclude that the space obtained from the $\SL$
action on such a vacuum character is infinite dimensional. This shows
that all theories with $c\geq 3$ are not rational.


\end{document}